\documentclass[english,twocolumn]{emulateapj}
\usepackage{epstopdf}

\usepackage{color}

\newcommand\lsim{\mathrel{\rlap{\lower4pt\hbox{\hskip1pt$\sim$}}
    \raise1pt\hbox{$<$}}}
\newcommand\gsim{\mathrel{\rlap{\lower4pt\hbox{\hskip1pt$\sim$}}
    \raise1pt\hbox{$>$}}} 
    \newcommand{\dm}{\mathrm {dm}}
\newcommand{\br}{\mathrm {b}}  
\newcommand{\vbc}{\mathrm{vbc}}
\newcommand{\bc}{\mathrm{bc}}
 
\newcommand{\tot}{\mathrm{tot}}

\usepackage{graphics}
\usepackage{amsmath}
\usepackage{graphicx}

\makeatother

\shorttitle{Stream velocity effects on gas fraction}
\shortauthors{Naoz et al.}
\makeatother

\begin{document}

\title{Simulations of Early Baryonic Structure Formation with Stream Velocity: II. The Gas Fraction}

\author{Smadar Naoz\altaffilmark{1,2}, Naoki Yoshida\altaffilmark{3}, Nickolay Y. Gnedin\altaffilmark{4,5,6}  }
\altaffiltext{1}{ Institute for Theory and Computation, Harvard-Smithsonian Center for Astrophysics, 60 Garden St.; Cambridge, MA, USA 02138}
\altaffiltext{2}{ CIERA, Northwestern University, Evanston, IL 60208, USA}
\altaffiltext{3}{Department of Physics, University of Tokyo, Tokyo 113-0033, Japan}
\altaffiltext{4}{Particle Astrophysics Center, Fermi National Accelerator Laboratory, Batavia, IL 60510, USA}
\altaffiltext{5}{Kavli Institute for Cosmological Physics and Enrico Fermi Institute, The University of Chicago, Chicago, IL 60637 USA}
\altaffiltext{6}{Department of Astronomy \& Astrophysics, The University of Chicago, Chicago, IL 60637 USA}
\email{snaoz@cfa.harvard.edu }

\begin{abstract}
Understanding the gas content of high redshift halos is crucial for
studying the formation of the first generation of galaxies and
reionization. Recently, Tseliakhovich \& Hirata showed that the
relative ``stream'' velocity between the dark matter and baryons at
the time of recombination - formally a second order effect, but an
unusually large one - can influence the later structure formation
history of the Universe. We quantify the effect of the stream velocity
on the so-called ``characteristic mass'' - the minimum mass of a dark
matter halo capable of retaining most of its baryons throughout its
formation epoch - using three different high-resolution sets of
cosmological simulations (with separate transfer functions for baryons
and dark matter) that vary in box size, particle number, and the value
of the relative velocity between the dark matter and baryons. In order
to understand this effect theoretically, we generalize the linear
theory filtering mass to properly account for the difference between
the dark matter and baryonic density fluctuation evolution induced by
the stream velocity. We show that the new filtering mass provides an
accurate estimate for the characteristic mass, while other
theoretical ansatzes for the characteristic mass are substantially
less precise.
\end{abstract}

\section{Introduction}

Gas rich dark matter halos in the early universe serve as a nurturing
ground for dwarf galaxies \citep[e.g.,][and references
therein]{Ricotti+02,Ric+02,BCL02,BCL99,Abel+02,NNB,Yoshida+06,Y08_firstS,Greif10,Clark11,BY11}. Their
properties are important to quantify, as they are responsible for metal
pollution and ionizing radiation at the onset of structure formation
\citep[e.g.,][]{Shapiro+04,Ciardi+06,Hoeft+06,Gnedin+08,Okamoto+08,Trenti+09}. More than that,
even if the smallest of gas rich halos are too small for efficiently cooling via
atomic hydrogen lines and may not host actual galaxies,
these ``mini-halos'' may produce a 21-cm signature in future radio observations
(\citet{Kuhlen,Shapiro+06,NB08} but see \citet{Furlanetto06}) and might
 block some of the ionizing radiation, causing an overall delay in the
initial progress of reionization \citep[e.g.,][]{BL02, iliev2, iss05,
mcquinn07}. Thus, the evolution of the gas fraction of dark matter halos at various epochs during the early evolution of the universe is of prime importance.

Recently, \citet{Tes+10a} showed that not only the amplitudes of the
dark matter and baryonic density fluctuations were different at early
times, but also were their velocities. After recombination, the sound
speed of the baryons dropped dramatically, while the dark matter velocity
remained high - thus, the relative velocity of baryons with respect to
the dark matter became supersonic. \citet{Tes+10a} also showed that
this relative velocity between the baryons and the dark matter remained
coherent on scales of a few mega-parsec and was of the order of $\sim
30$~km~sec$^{-1}$ at the time of recombination.  This relative
velocity is often called the ``stream velocity'' in the literature,
and throughout this paper we will use this term. The stream velocity
effect has previously been overlooked, because the velocity terms are
formally of the second order in the perturbation theory and should be
neglected in the linear approximation. However, this second order effect is
unusually large, resulting in the numerically non-negligible
suppression of power at mass scales that correspond to the
first bound objects in the Universe \citep[e.g.,][]{Yoshida+03early}.

Using the Press-Schechter \citep{ps} formalism, \citet{Tes+10a} showed
that the number density of halos is reduced by more than $60\%$ for
halos with $M=10^6$~M$_\odot$ at $z=40$.  In a subsequent paper,
\citet{Tes+10b} also included the baryonic temperature fluctuations  following
\citet{NB05}. They found that the stream velocity also resulted in
much higher ``characteristic'' mass - the minimum mass for a dark
matter halo capable of retaining most of its gas - as compared to the
case without the stream velocity \citep[e.g.][]{NB07}. As has been
shown in subsequent studies, the stream velocity effect has important
implications on the first structures
\citep{Stacy+10,Maio+11,Greif+11,Naoz+11a,Fialkov+11,OLMc12,BD} and
may also affect the redshifted cosmological 21-cm signal
\citep{Dalal+10,Bittner+11,Yoo+11,Visbal+12,McOL12}.

In this paper we explore the effect of stream velocity on the gas
fraction in dark matter halos and compare the simulation results to the
predictions from the linear theory \citep[e.g.][]{Tes+10b}. In our
first paper \citep[][hereafter Paper I]{Naoz+11a} we quantified the
stream velocity effect on the evolution of the halo mass function with
cosmological simulations. We used three different sets of high
resolution simulations in order to study the stream velocity effect
\emph{systematically}, thus understanding the overall trends (instead
of concentrating on specific halos). We used a set of simulations with
different box sizes, particle numbers, and the values for the stream
velocity to analyze the suppression of the structure formation as a
function of the stream velocity. In Paper I we found that  the
{\it total} number density of halos is suppressed by $\sim 20\%$ at
$z=25$ in regions of the universe that happen to have
$v_\bc=1\sigma_\vbc$, where $\sigma_\vbc$ is the (scale independent)
rms fluctuation of the stream velocity on small scales. In rare patches where
$v_\bc=3.4\sigma_\vbc$, the relative suppression at the same redshift
reaches $50\%$, remaining at or above the $30\%$ level all the way to
$z=11$.  Perhaps the most interesting phenomenon that we found was the
high abundance of ``empty halos'', i.e., halos that had their gas
fractions below half of the cosmic mean baryonic fraction
$\bar{f}_\br$. Specifically, we found that for $v_\bc=1\sigma_\vbc$
{\it all} halos below $10^5$~M$_\odot$ are empty at $z\geq 19$. As a
result, the high abundance of empty halos can significantly delay the
formation of gas rich ``minihalos'' and the first galaxies. In this
paper we investigate the effect of the stream velocity on the gas
fraction in halos. In particular, we quantify the dependence of the
characteristic mass on the magnitude of the stream velocity.

For completeness we first describe the parameters and initial
conditions of our simulations in \S \ref{sim}.  We present our results
and analysis of the gas fraction in halos and comparison to the linear
approximation in Section \ref{sec:McMf}.  Finally we offer a
brief discussion in \S \ref{sec:dis}.

Throughout this paper, we adopt the following cosmological
parameters: ($\Omega_\Lambda$, $\Omega_{\rm M}$, $\Omega_b$, n,
$\sigma_8$, $H_0$)= (0.72, 0.28, 0.046, 1, 0.82, 70 km s$^{-1}$
Mpc$^{-1}$)  \citep{wmap5}.

\section{The simulations}\label{sim}

\subsection{Basic Parameters and Settings}\label{sec:basic}

In this work we use a parallel $N$-body/hydrodynamics code GADGET-2
\citep{Gadget,G2}. Below we describe the general features of our 3
simulation sets, which are also summarized in table \ref{table_sim}.
\begin{enumerate}
\item The first set, named ``$N=256$'', uses a total of $2\times256^3$
  dark matter and gas particles within a cubic box 
  of $200$~comoving kpc on a side. To realize statistically significant 
  number of halos
  in such a small box, we artificially increase gravitational
  clustering in the simulation by setting $\sigma_8=1.4$.  We choose
  this box size so that a $10^4$~M$_{\odot}$ halo is resolved with
  $\sim500$ particles - the value needed to estimate the halo gas
  fraction reliably \citep{NBM}.  The gravitational softening is set to be 
  $40$~comoving~pc,
  well below the virial radius of a $10^4$~M$_\odot$ halo
  ($\sim 680$~comoving~pc). All the simulations in this set are initialized at
  $z=199$.
\item The second set, named ``$N=512$'', uses a total of
  $2\times512^3$ dark matter and gas particles within a cubic box with
  the size of $700$~kpc. In this set we also artificially increase
  $\sigma_8$ to 1.4. With these parameters, a halo with 500 dark
  matter particles has a mass of $\sim 5 \times 10^4$~M$_\odot$. The
  softening length is set to be $68$~comoving~pc. All the simulations in this set
  are initialized at $z=199$.
\item The final set of simulations uses $2\times768^3$ dark matter and
  gas particles (which we name the ``$N=768$'' set) in a $2$~Mpc box,
  and starts at $z=99$. For these parameters a halo with $500$ dark
  matter particle has a mass of $\sim 10^5$~M$_\odot$. The softening
  length is set to be $0.2$ comoving kpc. We use the ``correct'' value of
  $\sigma_8=0.82$ for this simulation set.
\end{enumerate}
In  each simulation set, we explore a range of the values for 
the stream velocity (see  table \ref{table_sim}). 

\begin{table}
 \caption{Parameters of the simulations}
\label{table_sim}
\begin{center}
\begin{tabular}{l c c }
\hline
{SIM}&  $v_{\rm bc,0}$ & $\sigma_\vbc$ stream       \\
  &   km~sec$^{-1}$&   velocity             \\
\hline \hline
  $256$ runs, & $0.2$~Mpc,& $z_{in}=199$\\
\hline \hline
256$_0$           & 0   & 0     \\
256$_{1\sigma}$            & 5.8   &  1       \\
256$_{1.7\sigma}$           & 10   &     1.7     \\
256$_{3.4\sigma}$            & 20  &  3.4      \\
\hline\hline
 $512$ runs, &$0.7$~Mpc, & $z_{in}=199$\\
\hline\hline
512$_0$     &0     & 0     \\
512$_{1\sigma}$     &   5.8   &1    \\
512$_{1.7\sigma}$     &   10 &1.7     \\
512$_{3.4\sigma}$     &   20 &3.4  \\
\hline\hline
 $768$ runs,&$2$~Mpc,& $z_{in}=99$\\
\hline\hline
768$_0$     &0     & 0    \\
768$_{1\sigma}$     &   3   &1    \\
768$_{1.7\sigma}$     &   5  &1.7     \\
768$_{3.4\sigma}$         & 10     &     3.4     \\
\hline
\vspace{-0.7cm}
\end{tabular}
\end{center}
\end{table}

\subsection{Initial Conditions}

As has been shown by \citet{NNB} and \citet{NB07}, setting up initial
conditions for cosmological simulations on small spatial scales is a delicate issue. High
accuracy in initial conditions is crucial for accurately predicting
the halo mass function in the lowest mass regime ($M \la
10^7$~M$_{\odot}$).

Following \citet{Naoz+10}, we generate separate transfer functions for
dark matter and baryons as described in \citet{NB05}. Ideally,
the stream velocities should be realized in the initial conditions in a
self-consistent way with the transfer functions that are calculated up
to the second order in the perturbation theory \citep[c.f. recent
studies by][]{OLMc12,McOL12}. However, that would require computing
transfer functions up to the second order in perturbation theory, and
those are not readily available. Instead, similar to all previous
simulation studies, we used the transfer functions computed in the
linear approximation. We account for that choice in \S \ref{sec:McMfcom}, when we compare our
simulations to the perturbations theory, and thus our results remain self-consistent.

For all runs, glass-like initial conditions were generated using
Zel'dovich approximation. For baryons, we have used a glass file with
positions shifted by a random vector, thus removing artificial
coupling between nearby dark matter and gas particles
\citep{Yoshida03b}.  We note that we have used the same phases for
dark matter and baryons in all of our simulations, since we showed in
Paper I that the spatial shift between baryons and dark matter is
unimportant.

\citet{Tes+10a} demonstrated that, while the stream velocity varies in
space, its coherence length is quite large, many Mpc. Hence, on scale
of our simulation boxes, it can be treated as constant bulk motion of
baryons with respect to the dark matter. We include the effect of stream
velocity by adding, at the initial redshift, an additional velocity to
the $x$ component of the baryons velocity vector. We test a range of
values for the stream velocity, which is convenient to quantify in
terms of its rms value on small scales, $\sigma_\vbc$. Specifically,
we test $v_\bc=1\sigma_\vbc$ through $v_\bc=3.4\sigma_\vbc$ for all
the simulations sets (see table \ref{table_sim}).

\subsection{Halo Definition}\label{sec:Hdef}

We locate dark matter halos by running a friends-of-friends group
finder algorithm with a linking parameter of $0.2$ (only for the dark
matter component). We use the identified particle groups to find the
center of mass of each halo.  After the center is located, we
calculate density profiles of dark matter and baryons separately,
assuming a spherical halo and using $2000$ radial bins between $r_{\rm
min}=0$~kpc and $r_{\rm max}=20$~kpc. Using the density profiles, we
find the virial radius $r_{vir}$ at which the total overdensity is
$200$ times the mean background density, and compute the mass and the
gas fraction of each halo within that radius. Recently, \citet{OLMc12}  used an unconventional definition for halos by using the highest {\it baryonic} density peaks as the center their halos. This method preferentially results in
larger gas fraction for $\sigma_\vbc\neq0$, compare to our more conservative method. 

Recently, \citet{More+11} showed that halos identified by the
friends-of-friends algorithm enclose an average overdensity that is
substantially larger than $200$, and its specific value depends on the
halo concentration. In our approach we use the friends-of-friends
algorithm only to find the center of mass of a halo, and compute
the actual halo mass using the spherical overdensity of $200$.

We only retain halos that contain at least $500$ dark matter particles
within their virial radii. The choice allows us to estimate halo
masses to about 15\% precision \citep{Trenti_halos} and to estimate
halo gas fractions reliably to a similar level of accuracy
\citep{NBM}. However, for some of our fit calculations we also include
halos with the number of particles as low as $100$; if we do that, we
assign a lower weight in the fit to these halos according to the
resolution study done in \cite{NBM}, see Appendix \ref{app:fit} for
more details.

\section{Results}\label{sec:McMf}

Let us consider the various scales involved in the formation of cosmic
structure. On large scales gravity dominates other forces and gas
pressure can be neglected. On small scales, on the other hand, the
pressure dominates gravity and prevents baryon density fluctuations
from growing together with the dark matter fluctuations. The relative
force balance at a given time can be characterized by the ``filtering
scale'' \citep{cs} - a physical scale above which a small gas
perturbation can grow due to gravity overcoming the pressure
gradient. In the non-evolving background, the filtering scale
coincides with the classical \citet{jeans} scale, but in the expanding
universe the two scales typically differ by a significant factor.

Immediately after recombination Compton scattering of Cosmic Microwave
Background (CMB) photons on the residual free electrons after cosmic
recombination kept the gas temperature coupled to that of the CMB, the
Jeans mass was constant in time and equal to the filtering
scale. However, at $z\sim 130$, the gas temperature decoupled from the
CMB temperature, the Jeans mass began to decrease with time as the gas
cooled adiabatically, and the filtering scale lagged behind the Jeans
scale.

Based on results from an early numerical simulations, \citet{gnedin00}
suggested that the filtering mass also describes the largest halo mass
whose gas content is significantly suppressed compared to the cosmic
baryon fraction. The latter mass scale, commonly called the
``characteristic mass'', is defined as the halo mass for which the
enclosed baryon fraction equals half of the cosmic mean.
Thus, the characteristic mass distinguishes between gas-rich and
gas-poor halos. Many semi-analytical models of dwarfs galaxies 
use the characteristic mass scale in order to estimate the gas
fraction in halos
\citep[e.g.,][]{Bullock,Benson02a,Benson02b,Somerville,BD}. Theoretically
this sets an approximate minimum value on the mass that can still form
stars.

\subsection{Non-linear Behavior: the Characteristic Mass }\label{sec:Mc}

 For halos, \citet{gnedin00} defined a characteristic mass
$M_c$ for which a halo contains half the mean cosmic baryon fraction
$f_b$. In his simulation he found the mean gas fraction in halos of a
given total mass $M$, and fitted the simulation results to the
following formula:
\begin{equation}
\label{f_g-alpha}
f_{g,\rm calc}= f_{\br,0} \bigg[1+\left(2^{\alpha/3}-1
\right)\left(\frac{M_c}{M}\right)^\alpha \bigg]^{-3/\alpha} \ ,
\end{equation}
where $f_{\br,0}$ is the gas fraction in the high-mass limit\footnote{Following \citet{Naoz+10} we defined the high-mass bin as the largest $5\%$ halo mass, or the largest $5$ halos (if the larges $5\%$ consists of less than $5$ halos). These halos are usually gas rich and if when we miss them it causes
 to an underestimate of the gas fraction at the larger mass bin. \citet{Naoz+10} found this method  to be consistent with setting a low linking parameter, and with varying  the halo radii between $r_{100}$, $r_{200}$ and $r_{500}$. This way we overcome the disadvantages of assuming a spherical halo, which misses gas rich halos which undergoes mergers. }.  In this
function, a higher $\alpha$ causes a sharper transition between the
high-mass (constant $f_g$) limit and the low-mass limit (assumed to be
$f_g \propto M^3$). \citet{gnedin00} found a good fit for $\alpha =
1$, with a characteristic mass that in fact equaled the filtering mass
by his definition. 

\begin{figure}[!t]
  \centering \includegraphics[clip,scale=0.4]{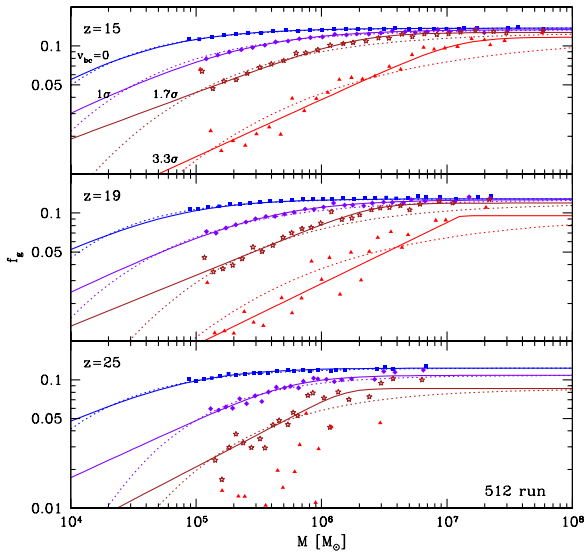}
\caption{Binned gas fraction in the ``N=512'' runs at redshifts 25, 19, and
  15 (from bottom to top panels) for 4 values of the stream velocity
  $v_\bc=0$, $v_\bc=1\sigma_\vbc$, $v_\bc=1.7\sigma_\vbc$, and
  $v_\bc=3.4\sigma_\vbc$ (blue squares, purple diamonds, brown stars
  and red triangles, respectively). We also show the best fit as
  evaluated from equations (\ref{f_g-alpha}) and (\ref{f_g-new}),
  with dotted and solid lines respectively.  Note for the case of $v_\bc=3.4\sigma_\vbc$,  at $z=25$, about $2/3$ of the halos have less than $1 \%$ of gas in them, and $f_{\br,0}$ is very low. Thus no reliable  fit could be found, and therefore we did not show the resulted red line in that case.  (See also Figure
  \ref{fig:fit_fg} for the gas fraction behavior for all halos, 
  i.e., not binned.)    }
\label{fig:fg512}
\end{figure}

The characteristic mass is essentially a non-linear version of the
filtering mass, and so it also measures the competition between
gravity and pressure. At high masses, where pressure is unimportant,
$f_g\to f_{b,0}$, while the low mass tail is determined by the
suppression of gas accretion by gas pressure.  \citet{NBM} found that
the filtering mass from linear theory (calculated in a self consistent
way) is consistent with the characteristic mass fitted from the
simulations, for two (pre-reionization) scenarios that they tested: a
case with no stellar heating and a case of a sudden flash of stellar
heating at a given redshift. In a followup paper, \citet{Naoz+10} found the same agreement between the 
the linear and non-linear theory, and showed that alternative initial conditions models yield a different (higher by about 50 per cent) minimum mass (both the linear and non-linear), since the system retains a memory of the initial conditions. 
For clarity, we emphasize that the
statement ($M_c=M_F$) refers to our definition of $M_F$ in
equation~(\ref{Mf}).

In Figure \ref{fig:fg512} we present the gas fraction as a function of
halo mass for our ``N=512'' simulation set\footnote{In Paper I we
showed that, qualitatively, all of our simulation sets behave
similarly as a function of mass and redshift. Therefore, to avoid
redundancy, we show here the gas fractions for only ``N=512'' runs.}. As
one can see, the halo gas fraction drops dramatically at lower halo
masses for large values of the stream velocity. This trend introduces
a qualitatively different behavior in the gas fraction as a function
of halo mass, which is not captured by equation (\ref{f_g-alpha}).  The
best fit ansatz [Eq.~(\ref{f_g-alpha})] is shown with dotted lines in Fig.\
\ref{fig:fg512}. It clearly does not capture the behavior of the gas
fraction as a function of halo mass for $v_\bc\neq 0$. Therefore we
introduce a new fitting formula for the gas fraction as a function of
halo mass,
\begin{equation}
\label{f_g-new}
f_{g,\rm calc}= f_{\br,0} \bigg[1+\left(2^{\gamma}-1
\right)\left(\frac{M_c}{M}\right)^\beta \bigg]^{-1/\gamma} \ .
\end{equation}
The new fitting formula reduces to Equation (\ref{f_g-alpha}) for
$\gamma = 3\beta = \alpha$.

Although we add another free parameter to the fit, finding the best
fit model presents some hurdles which we discuss in the Appendix. The
best-fit value of $M_c$ from Equation (\ref{f_g-new}) is the same as
the value from Equation (\ref{f_g-alpha}) for the $v_\bc=0$ case, as
can be expected from the fact that Equation (\ref{f_g-alpha}) provides
a good fit to the simulation results in the $v_\bc=0$ case.  It is
interesting, however, that the best fit values for $M_c$ from Equation
(\ref{f_g-new}) are consistent with the best fit values of   $M_c$ from Equation (\ref{f_g-alpha}),  even
for large values of $v_\bc$, as can be seen in Figure
\ref{fig:fit_calc}. However, the new fitting formula gives a better
overall fit for high $v_\bc$ values, especially in the high mass limit.

\begin{figure}[!h]
  \centering \includegraphics[clip,scale=0.4]{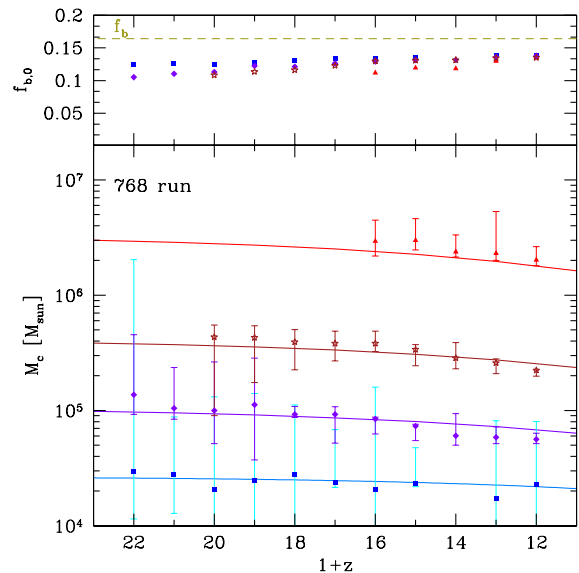}
\caption{Best fit values for the characteristic mass and limiting
 baryon fraction $f_{\br,0}$ as functions of redshift for various
 values of the stream velocity $v_\bc=0$, $v_\bc=1\sigma_\vbc$,
 $v_\bc=1.7\sigma_\vbc$, and $v_\bc=3.4\sigma_\vbc$ (blue squares,
 purple diamonds, brown stars and red triangles respectively) for our
 $N=762$ simulation set. We also show the evolution of the fully
 self-consistent filtering mass (Equation \ref{kfnew}) with solid
 curves.  The error bars are the maximum $1-\sigma$ from the two fit models (see Appendix \ref{app:fit}). }
\label{fig:Mc768}
\end{figure}

\begin{figure}[!h]
  \centering \includegraphics[clip,scale=0.4]{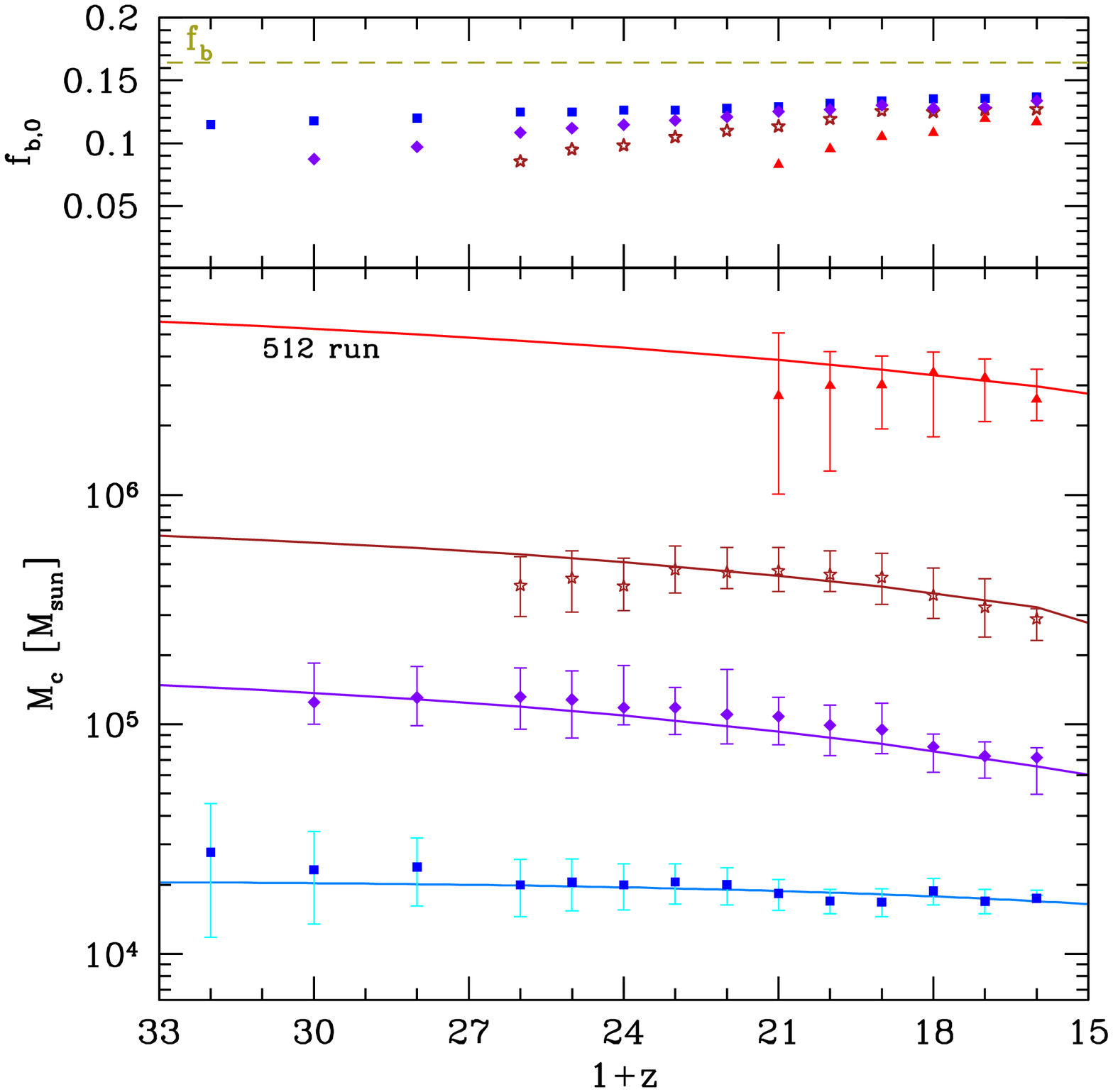}
\caption{Same as Figure \ref{fig:Mc768}, but for the $N=512$ simulation set.}
\label{fig:Mc512}
\end{figure}

\begin{figure}[!h]
  \centering \includegraphics[clip,scale=0.4]{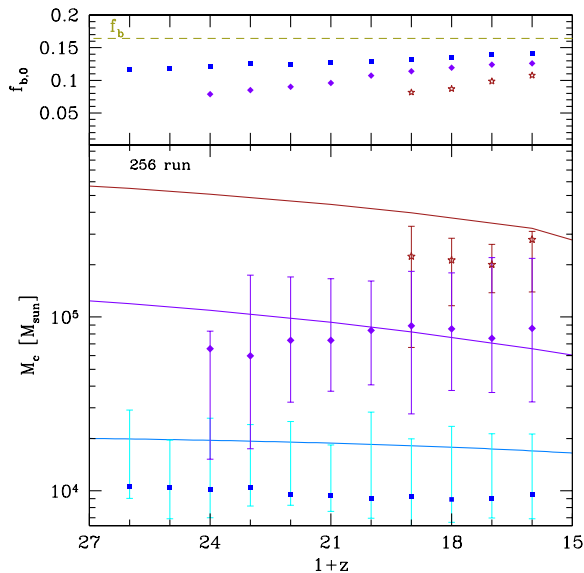}
\caption{Same as Figure \ref{fig:Mc768}, but for the $N=256$ simulation set.}
\label{fig:Mc256}
\end{figure}

Figures \ref{fig:Mc768}--\ref{fig:Mc256} show the best-fit values for
the characteristic mass and limiting baryon fraction $f_{\br,0}$ at a
range of redshifts. At the highest redshifts and large values of
$\sigma_\vbc$ most of the halos are empty halos (i.e., halos with gas
fraction lower then the half of the mean cosmic baryonic fraction, see
Paper I, figures 2-4). This of course means that  there is no apparent trend similar to Figure \ref{fig:fg512}, and  therefore, no convergence could be achieved in
the fitting procedure and the parameters of the fit cannot be
measured. Furthermore, as has been noted in Paper I, the $N=256$ set
suffers from poor statistics for $M \ga 10^5{\rm M}_\odot$, resulting
in large error bars for the best-fit values of the parameters.  We show the results of this run 
to caution the interpretations of previous (and perhaps some future) which employ very small boxes.

An important point to make is that at low redshifts the baryon fraction at the highest mass bins (for all runs) approaches the same value irrespectively of the magnitude of the stream velocity. This is, of course, expected, as the global trend for the baryon fraction is to approach the cosmic mean (see Appendix \ref{app:fit}, Figure \ref{fig:fg_fbbar}), but still below it at high redshift (even at large scales). This is because the baryons still did not fall into the Dark Matter potential wells \citep[see][and the Appendix for further discussion]{NB07,BL11}.  However, in the case of the largest stream velocity we consider, $v_\bc=3.4\sigma_\vbc$, the $N=256$ and $N=512$ simulation sets do not quite reach the cosmic mean values expected. This is most likely due to the low abundance of the most massive halos, since some baryons are in fact bound to the halo and we miss them due to our  halo finder algorithm (\S \ref{sec:Hdef}), as can be seen in Figure 4 of Paper I. The $N=768$ simulation set, however, does not suffer from that incomplete convergence.
We find a simple fit for the gas fraction in the last mass bin, $f_{\br,0}$, for the {\it low redshift} limit as a function of the stream velocity 
\begin{equation}
f_{\br,0}= -0.0049\frac{v_\bc}{\sigma_\vbc}+0.1345 \ .
\end{equation}

Note that throughout the paper we compare between different simulations using different $\sigma_8$ values (i.e., $\sigma_8=0.82$ for the $N=768$ set and $\sigma_8=1.4$ for the $N=512$ and $N=256$ sets). As was shown in Paper I the  suppression of  the halo mass function, due to the stream velocity, compare to the no stream velocity case is independent on $\sigma_8$ (see Figures 5 and 6 in Paper I). Since the characteristic mass describes a relative suppression of the gas fraction in small scales compare to large scales, we find that this quantity is independent on $\sigma_8$ as well (where the different simulations gave a consistent value of $M_c$). This is further supported by the agreement to linear theory (see below). Thus, increasing $\sigma_8$
only raised the clustering amplitude that enlarges the sample of simulated halos.

\subsection{Linear Theory Predictions: the Filtering Mass}\label{sec:Mf}

In the linear approximation, the filtering mass, first defined by
\citet{cs}, describes the highest mass scale on which the baryon
density fluctuations are suppressed significantly compared to the dark
matter fluctuations. \cite{cs} only considered the low redshift case,
where the baryonic and dark matter fluctuations have the same
amplitude at large scales. \citet{NB07} relaxed that assumption and
extended the computation of the filtering mass to early times, during
which the amplitude of the baryonic fluctuations is below the
amplitude of the dark matter fluctuations even on large scales. Both
studies, however, only considered a case of zero stream velocity.

In order to extend the derivation of the filtering mass to the non-zero stream velocity case, we first introduce the coupled
second order differential equations that govern the evolution of the
density fluctuations of the dark matter ($\delta_\dm$), and the
baryons ($\delta_\br$) and the baryon temperature ($\delta_T$):
\begin{eqnarray}\label{g_T}
\ddot{\delta}_{\dm} + 2H \dot {\delta}_{\dm}- f_\dm\frac{2 i}{a}  {\bf v}_\bc \cdot {\bf k} \dot\delta_\dm  & = &
 \\
\frac{3}{2}H_0^2\frac{\Omega_{m}}{a^3}
\left(f_{\br} \delta_{\br} + f_{\dm} \delta_{\dm}\right)& +& \left( \frac{  {\bf v}_\bc \cdot {\bf k}} {a} \right)^2 \delta_\dm   \nonumber  \\
\ddot{\delta}_{\br}+ 2H \dot {\delta}_{\br}  \ \ \ \ \ \ \ \ \ & = &
  \\
\frac{3}{2}H_0^2\frac{\Omega_{m}}{a^3} \left(f_{\br}
\delta_{\br} + f_{\dm}
\delta_{\dm}\right) &-& \frac{k^2}{a^2}\frac{k_B\bar{T}}{\mu}
\left(\delta_{\br}+\delta_{T}\right) \ , \nonumber \label{eq:db}
 \end{eqnarray}
where $\Omega_m$ is the present day matter density as a fraction of the
critical density, $k$ is the comoving wavenumber, $a$ is the scale
factor, $\mu$ is the mean molecular weight, $H_0$ is the present day
value of the Hubble parameter $H$, and $\bar{T}$ and $\delta_T$ are the
mean baryon temperature and its dimensionless fluctuation,
respectively. These equations are a compact form of equations 5 in \citet{Tes+10a}, where we used the fact that $v_\bc \propto 1/a$, and 
the baryon equation includes
the pressure term whose form comes from the equation of state of an
ideal gas. The linear evolution of the temperature fluctuations is
given by \citep{BL05,NB05}
\begin{equation}
\label{gamma} \frac{d \delta_T} {d t} = \frac{2}{3} \frac{d
\delta_\br} {dt} + \frac{x_e(t)} {t_\gamma}a^{-4} \left\{
\delta_\gamma\left( \frac{\bar{T}_\gamma}{\bar{T}} -1\right)
+\frac{\bar{T}_\gamma} {\bar{T}} \left(\delta_{T_\gamma} -\delta_T
\right) \right\}\ ,
\end{equation}
where $x_e(t)$ is the free electron fraction as a function of cosmic
time $t$, $\delta_\gamma$ is the photon density fluctuation,
$t_\gamma=8.55 \times 10^{-13} {\mathrm{yr}}^{-1}$, and $T_\gamma$ and
$\delta_{T_\gamma}$ are the mean photon temperature and its
dimensionless fluctuation, respectively. Equation~(\ref{gamma})
describes the evolution of the gas temperature in the
post-recombination era, but before formation of first galaxies, when
the only external heating arises from Compton scattering of the
remaining free electrons on the CMB photons. The first term in
Equation~(\ref{gamma}) comes from the adiabatic cooling or heating of
the gas, while the second term is the result of the Compton
interaction.

\begin{figure*}[htb]
  \centering \includegraphics[width=12cm,clip=true]{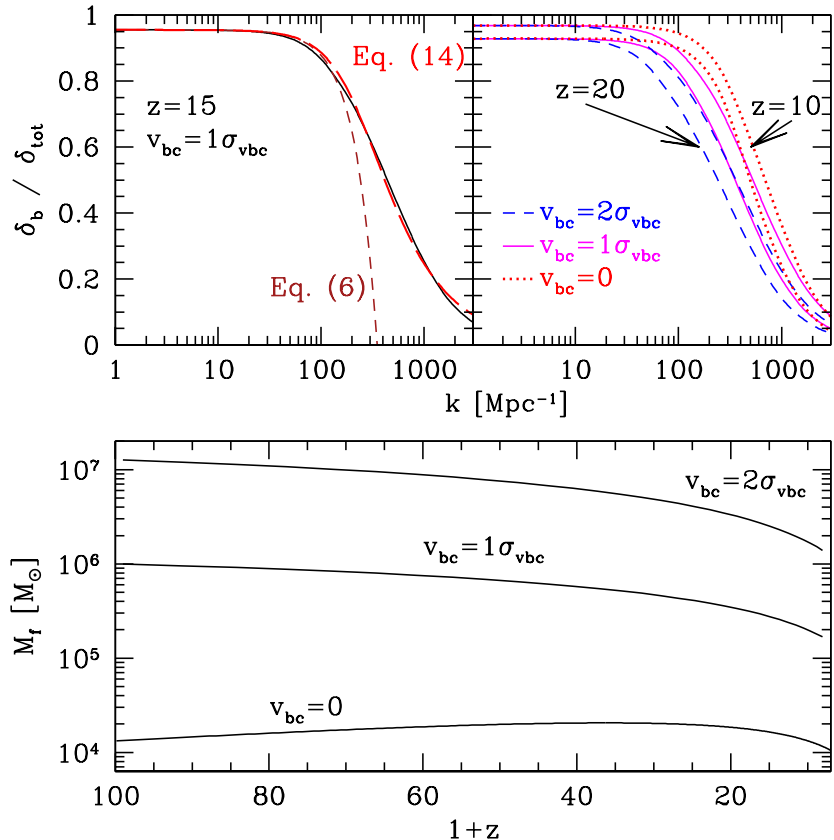}
\caption{ Fully self-consistent linear calculation of the growth of the dark matter and baryonic fluctuations \citep[i.e., following][]{Tes+10a}. We show in the bottom panel the filtering mass as a function of redshift in the regions with $v_\bc=0,1$ and $2$~$\sigma_\vbc$ (see labels). In the top panels we show $\delta_{\br}/\delta_\tot$ as a function of the wave number $k$. In the right top panel we plot  $\delta_{\br}/\delta_\tot$ in the regions with $v_\bc=0$ (dotted red lines),$1\sigma_\vbc$ (solid magenta lines) and $2\sigma_\vbc$ (dashed blue lines) at $z=10$ (top lines) and $z=25$ (bottom lines), see labels. In the top left panel we show an example for the $v_\bc=1\sigma_\vbc$ case at $z=15$ and plot the two fitting functions (see text for details), one which reproduces the drop of  $\delta_{\br}/\delta_\tot$ with wavenumber, Eq.~(\ref{fit}) (red dashed line) and the fitting function to the second order, i.e., Eq.~(\ref{kfnew}), (brown short dashed line).
 }
\label{fig:LinMf}
\end{figure*}

In the top right panel of Figure \ref{fig:LinMf} we show an example of
the solution of Equations (\ref{g_T}). We plot the ratio for
$\delta_\br / \delta_\tot$ as a function of the wavenumber $k$ for the
fully self-consistent linear calculation, i.e., starting at the time
of recombination and using the exact transfer functions from
\citet{Tes+10a}. We consider cases with $v_\bc=0$,
$v_\bc=\sigma_\vbc$, and $v_\bc=2\sigma_\vbc$ at $z=10$ and
$z=25$. For larger $v_\bc$ values, the drop in $\delta_\br /
\delta_\tot$ occurs at larger scales, i.e., the suppression of the
baryonic perturbations relative to the total matter fluctuations shifts
to larger masses.

 \citet{NBM} and \citet{Naoz+10},  showed that the characteristic mass is in a good agreement with the filtering mass, regardless of the initial conditions, or even if heating is involved, as long as the filtering mass is calculated self consistently. Motivated by these results we set to find a filtering mass that can be calculated self consistently and that will present the excepted agreement with the simulations.
Following \citet{NB07}, we re-define the filtering scale
(specifically, the filtering wavenumber $k_F$) to include the stream
velocity effect as
 \begin{equation}
\frac{\delta_\br}{\delta_\tot}=1+r_{\rm LSS}-\frac{k^2}{k_F^2} \frac{1}{1+\nu} \ ,
\label{kfnew}
\end{equation}
where $\nu={v_\bc}/{\sigma_\vbc}$ and $\sigma_\vbc$ is the
(scale-independent) rms of the stream velocity at small scales.
The parameter $r_{\rm LSS}$ (a
negative quantity) describes the relative difference between
$\delta_\br$ and $\delta_\tot$ on {\em large scales}
\citep{NB07}, i.e.,
\begin{equation}
r_{\rm LSS} \equiv \frac{\Delta}{\delta_\tot}\ ,
\label{r_LSS}
\end{equation}
where $\Delta=\delta_\br-\delta_\mathrm{tot}$ \citep[see
also][]{BL05}.
The filtering mass is
defined from $k_F$ simply as:
\begin{equation}
M_F=\frac{4\pi}{3}\bar{\rho_0}\left(\frac{1
}{2}\frac{2\pi}{k_F}\right)^3\ , \label{Mf}
\end{equation}
where $\bar{\rho_0}$ is the mean matter density today. 

To find  $k_F$  in a general case, we write it in the form
\begin{equation}
k_F^2(t)=\frac{\delta_\tot}{ u(t)} \ ,
\end{equation}
where $u(t)$ is to be determined.
Then, using equation (\ref{kfnew}), we expand the baryonic fluctuation as a function of wavenumber $k$,
\begin{equation}\label{eq:expand}
\delta_\br=\delta_\tot+\Delta_{\rm LSS}-\frac{u(t)k^2}{1+\nu} + O(k^4),
\end{equation}
where $\Delta_{\rm LSS} \equiv r_{\rm LSS}\delta_\tot$ [eq.~(\ref{r_LSS})] obeys the following equation to the first order of $k$,
\begin{equation}\label{eq:DeltaLS}
 \ddot{\Delta}_{\rm LSS}+2H\dot{\Delta}_{\rm LSS}= - \frac{2 i}{a} f_\dm  {\bf v}_\bc \cdot {\bf k} \dot\delta_\dm . 
 \end{equation}
Note that in the case of $v_\bc=0$, the linear term of $k$ has a zero coefficient, and thus the right hand side of this equation is simply zero \citep[see][]{BL05,NB07}.
Substituting the expansion from equation (\ref{eq:expand}) into equation (\ref{eq:db}), and using equations (\ref{g_T}) and (\ref{eq:DeltaLS}),  we obtain an equation for
$u$:
  \begin{eqnarray}
  \label{ratio1}
\ddot{u}+2H\dot{u}&=& f_\dm \left( 1+\nu \right) \bigg \{ \frac{1}{a^2}\frac{k_B\bar{T}}{\mu}
\left(\delta_\br+\delta_T\right) \\
&+& \left( \frac{  {\bf v}_\bc \cdot {\bf k}} {a} \right)^2 \delta_\dm \bigg \} \ . \nonumber 
\end{eqnarray}
In the limit of $v_\bc=0$ (i.e., $\nu=0$) this equation reduces to equation (12) of \citet{NB07}, and thus results in the same filtering mass found in that study.
We can solve
Equation (\ref{ratio1}) to find $u(t)$,
\begin{eqnarray}
\label{u_eq}
\lefteqn{
u(t)= } \\ && f_\dm \left(1+ \nu \right) \bigg \{  \int^{t}_{t_{\rm rec}}\frac{dt^{\prime\prime}}{a^2(t^{\prime\prime})}
\int_{t_{\rm rec}}^{t^{\prime\prime}} dt^{\prime} \frac{k_B\bar{T}(t^\prime)}{\mu}\left(\delta_\br(t^\prime)+\delta_T(t^\prime)\right) \nonumber  \\  
&& + \int^{t}_{t_{\rm rec}}\frac{dt^{\prime\prime}}{a^4(t^{\prime\prime})}
\int_{t_{\rm rec}}^{t^{\prime\prime}} dt^{\prime} \left(  {\bf v}_{\bc,{\rm rec}}(t^\prime) \cdot \hat{\bf k} \right)^2 \delta_\dm(t^\prime) \bigg \}  \ . \nonumber
\end{eqnarray}
where ${\bf v}_{\bc,{\rm rec}}$ is the stream velocity at the moment of recombination $t_{\rm rec}$, so that ${\bf v}_\bc(t)={\bf v}_{\bc,{\rm rec}} a(t_{\rm rec}) / a(t)$, and $\hat{\bf k}={\bf k} /k$ is the unit wavenumber vector.
In the bottom panel of Figure \ref{fig:LinMf} we show the evolution of the filtering mass as a function of redshift for $v_\bc=0,1$ and $2\sigma_\vbc$. The values of the filtering mass for $v_\bc \neq 0$ as defined by Equation (\ref{kfnew}) are larger up to an order of magnitude at high redshifts as compared to the definition of \citet{Tes+10b}. We emphasize that this difference is entirely due to the different definition of the filtering scale, not due to any error in \citet{Tes+10b} calculations.

The filtering scale $k_F$  can also be obtained simply by fitting equation (\ref{kfnew}) to the calculated values of $\delta_\dm$ and $\delta_b$, using equations (\ref{g_T})--(\ref{gamma}).  \citet{NB07} found a 
functional form that can be used to produce a good fit for the drop of the wavenumber. Generalizing it to the case of stream velocity we write: 
\begin{equation}\label{fit}
\frac{\delta_\br}{\delta_\tot} \approx (1+r_{\rm LSS})\left(1+\frac{1}{n}
\frac{k^2/k_F^2}{1+r_{\rm LSS}}\frac{1}{1+\nu}\right)^{-n}\ ,
\end{equation}
and $n$ must be adjusted at each
redshift. 
In the top left panel of Figure \ref{fig:LinMf} we compare this fitting formula to the fully self-consistent linear calculation (long dashed line) for which reproduce the drop of ${\delta_\br} /{\delta_\tot}$ as a function of $k$ fairly well. For the example considered in the figure, i.e., $z=15$ and $v_\bc=1\sigma_\vbc$, we find $n=0.46$ and  $k_F=253.9$~Mpc$^{-1}$.  We also show the resulted fit using the second order in $k$ approximation, i.e.\ Eq.~(\ref{kfnew}).

\subsection{Comparison Between the Linear Theory Predictions and the Nonlinear Results}\label{sec:McMfcom}

In order to compare the filtering mass to the characteristic mass we calculated the filtering mass in a self consistent way, as was done in \citet{NBM,Naoz+10}. In other words we use the transfer function from \citet{NB05} with a boosted velocity for the baryons at $z=99$ ($z=199$) for the $N=768$ ($N=512$ and $N=256$) set as initial conditions.  We then evolve the dark matter and baryon in time according to equations (\ref{g_T})--(\ref{gamma}).   We note that in all our calculations we included the fact that the boost of the velocity was included in the simulation only in one axis, thus terms which are proportional to ${\bf v}_\bc \cdot {\bf k}$ are reduced by a factor $3$ compare to the global average. 

Our simulation sets $N=256$ and $N=512$ are initialized at redshift
$199$ at which Compton heating by the CMB photons significantly
affects the evolution of the linear modes and specifically the
filtering mass \citep{NB05,NB07}.  However, GADGET-2 does not include
CMB Compton heating. Hence, to compare apples and apples, we
neglected the Compton heating contribution to the filtering mass when comparing these two simulation sets to the linear approximation. This is the reason that our values
of the filtering mass in Figures \ref{fig:Mc512} and
\ref{fig:Mc256} are lower than the values for the no stream velocity case in Figure 3 of
\citet{NB07}.

We show the linear theory filtering mass as a function of redshift for all the cases we consider in Figures \ref{fig:Mc768}--\ref{fig:Mc256}. The filtering mass is consistent with the characteristic mass within our fit errors for all simulation sets. 
Therefore, we conclude in agreement with \citet{NBM,Naoz+10}, that the evolution of the characteristic mass can be understood using the linear approximation predictions for the filtering mass.

Note that we compare between the filtering mass and $M_c$ results from simulation runs using different $\sigma_8$. This is possible since the filtering  mass definition is independent on $\sigma_8$ because it describes the  ratio between the two density fluctuations (thus the normalization of the powers simply cancels out). 

\subsection{Comparison with  the Previous Definition of the Filtering Mass and Other Mass Scales}\label{sec:Mfold}

\begin{figure}[t]
  \centering \includegraphics[clip,scale=0.4]{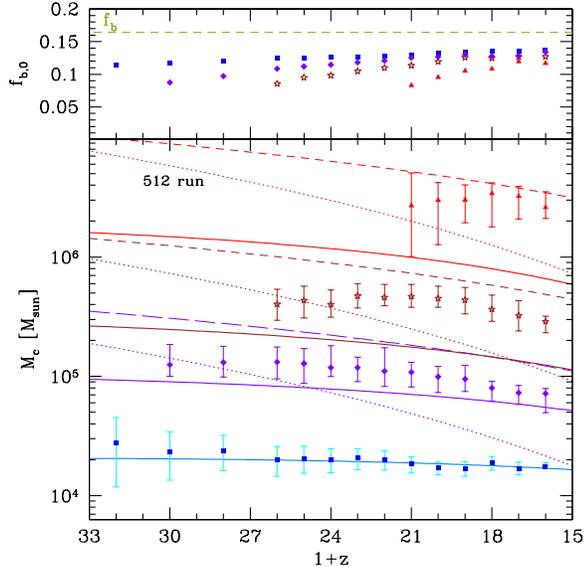}
\caption{Best fit values for the characteristic mass and limiting
 baryon fraction $f_{\br,0}$ as functions of redshift (same as in Fig.\ \ref{fig:Mc512}), but now compared with previous theoretical models for the characteristic mass. Solid lines show \citet{NB07} definition of the filtering mass (Eq.~\ref{kf_btot}) that neglects the stream velocity contribution. We also plot $M_{eff}$ (dashed lines) and $M_{esc}$ (dotted lines) as alternative models, see text for details.}
\label{fig:Mc512OLD}
\end{figure}

Recently \citet{Tes+10b}  showed that including the effects of the relative velocity between the dark matter and the baryons at recombinations results in a higher filtering mass as compared to the case of $v_\bc=0$ (by about order of magnitude for the global average).
\citet{NB07} defined the filtering scale in the case of $v_\bc=0$ as:
\begin{equation}
\frac{\delta_\br}{\delta_\tot}=1-\frac{k^2}{k_F^2}+r_{\rm LSS}\ .
\label{kf_btot}
\end{equation}
In Figure \ref{fig:Mc512OLD} we show thus defined filtering mass for the $N=512$ simulation set. As can be seen, the \citet{NB07} definition underestimates the characteristic mass in the high $v_\bc$ limit, since it neglects the difference between the dark matter and baryons density fluctuations. 

Recently \citet{Stacy+10} and \citet{Naiman+11} suggested that, given a high initial stream velocity, the baryon evolution is dominated by the relative motion of dark matter and gas, and thus the gas sound speed should be replaced with the effective sound speed $v_{\rm eff}$, 
\begin{equation}\label{eq:veff}
v_{\rm eff}=\sqrt{c_s^2(z)+v^2_{\rm bc}(z)} \ ,
\end{equation}
where $v_{\rm bc}(z)=v_{\rm bc,0} / (1+z)$, in the Jeans mass definition. This effective Jeans scale $k_{J,{\rm eff}}$ can be written as
\begin{equation}
k_{J,{\rm eff}}=\frac{a}{v_{\rm eff}}\sqrt{4\pi G\bar{\rho}_m} \ ,
\end{equation}
where $G$ is the gravitational constant and $\bar{\rho}_m$ is the average density. 
The effective Jean mass $M_{\rm eff}$ associated with this scale length is simply 
\begin{equation}
M_{\rm eff}=\frac{4\pi}{3}\bar{\rho_0}\left(\frac{1
}{2}\frac{2\pi}{k_{J,{\rm eff}}}\right)^3\ . \label{Meff}
\end{equation}
We show this mass scale in figures  \ref{fig:Mc512OLD} with dashed lines. As one can see, this mass scale overestimates the characteristic mass at all redshifts and for all values of the stream velocity that we considered. This is not unexpected, since the Jeans mass always overestimates the scale at which pressure starts to overtake gravity in the expanding background \citep{cs}.

The evolution of the characteristic mass as a function of redshift can also be modeled by considering the escape velocity of the gas. Given a mass of a halo $M$, the escape velocity is simply $v_{esc}=\sqrt{2 G M/r}$, where $r$ is the (comoving) virial radius of the halo. 
 For example, for a halo of $10^5$~M$_\odot$ the escape velocity is about
$0.77$~km~sec$^{-1}$, while the stream velocity for
$v_\bc=3.4\sigma_\vbc$ at $z=15$ is $1.6$~km~sec$^{-1}$ and at $z=25$
is $2.6$~km~sec$^{-1}$. Thus, it is not surprising that halos below
$10^5$~M$_\odot$ are empty in that redshift range - the stream
velocity is simply much larger then the halo escape velocity, so the
dark matter halo is unable to accrete any gas. 
We can, thus, estimate a halo mass $M_{\rm esc}$ below which the stream velocity is larger then the escape velocity,
  \begin{equation}\label{eq:vesc}
M_{\rm esc}=\frac{v_\bc(z)^3}{\sqrt{(2 G H_0)^2 \Omega_m \Delta_c}} \ ,
\end{equation}
where  $\Delta_c=200$ is the virial overdensity and $v_\bc(z)$ is the stream velocity at redshift $z$. We show this limit 
in Figure \ref{fig:Mc512OLD} with dotted lines. Equation (\ref{eq:vesc}) predicts a much stronger evolution of the characteristic mass than is actually observed in our simulations and provides a poor fit to simulations results.

\section{Conclusions}\label{sec:dis}

We have used three-dimensional hydrodynamical simulations to
investigate the effects of stream velocity on the gas fraction in high
redshift halos. In a companion paper \citet{Naoz+11a}, we studied the
effect of the stream velocity on the {\it total} halo mass function,
In this work we focus on the effect of the stream velocity on the gas fraction in halos and on the evolution of the characteristic mass, and compare the simulation results to the linear approximation.

In a first improvement over the earlier results, we introduce a new fitting formula (Eq. \ref{f_g-new}) which offers a much better fit to the gas fraction as a function of halo mass at a given redshift in the limit of large stream velocities, while returning essentially the same values of the characteristic mass $M_c$ as the previously used functional form (see Figures \ref{fig:fg512}, \ref{fig:fit_fg} and \ref{fig:fit_calc}, and see Appendix \ref{app:fit}). 

Previous studies \citep{NBM,Naoz+10} showed that a quantity defined in the linear approximation, the filtering scale \citep{cs}, provides a good match to the nonlinear characteristic mass measured in numerical simulations. We introduce a new definition for the linear filtering mass that accounts for two effects neglected in \citet{cs}: the deviation of the amplitude of baryonic fluctuations from the dark matter fluctuations on large scales \citep[considered first by][]{NB07} and the stream velocity between the dark matter and baryons on small scales, which we include in the definition of the filtering mass for the first time in this paper. The latter effect may result in the filtering mass being up to an order of magnitude larger at high redshifts for high values of the stream velocity, as compared to the case when the stream velocity is neglected.

Finally, in comparing our simulations results to the linear calculation (using our new definition of the filtering mass), we find that the filtering mass (i.e.\ a linear quantity) offers an accurate match to the actual nonlinear characteristic mass measured from the simulations, at all redshifts and for all values of the stream velocity that we simulated. On the contrary, previous theoretical models that used as the characteristic mass scale either the halo mass with the escape velocity equal to the stream velocity or the Jeans mass for the ``effective'' gas sound speed provide only poor fits to the simulation results.

It has been suggested in the literature that gas rich low mass halos
may play an important role in cosmic reionization, and that they can
produce distinct 21-cm signatures (\citet{Kuhlen,Shapiro+06,NB08} but
see \citet{Furlanetto06}). For example, minihalos (halos of mass
$\sim10^6{\rm M}_\odot$) can potentially block ionizing radiation and
induce an overall delay in the initial progress of reionization
\citep[e.g.,][]{shapiro87,BL02,iliev+03,Shapiro+04,iss05,
mcquinn07}. However, our results here suggest that at high redshifts
the stream velocity effect results in large variations in the
characteristic mass - i.e. the minimum mass of a gas rich halo. Thus,
if reionization started sufficiently early \citep{Yoshida+07}, in
patches of the universe where the stream velocity is large there were
fewer gas rich halos that can absorb ionizing photons.  Hence, in
these patches the delay of the reionization caused by minihalos would
be less than in regions that happen to have a small value of the
stream velocity and, hence, a large abundance of minihalos. Therefore,
not only the formation of the first generations of galaxies may be
affected by the stream velocity effect, but also the whole process of
reionization may proceed differently in regions with very different
stream velocities. This effect has been considered recently by
\citet{Visbal+12} and \citet{McOL12}, but our results indicate that it
can even be stronger than previously estimated.

\section*{Acknowledgments}
We thank Avi Loeb, Rennan Barkana, Andrey Kravtsov, Neal Dalal, Will
Farr, Matt  McQuinn and Dmitriy Tseliakhovich for useful discussions.  We  thank
Dmitriy Tseliakhovich for providing his code. We also thank Yoram
Lithwick for the use of his allocation time on the computer cluster
Quest. This research was supported in part through the computational
resources and staff contributions provided by Information Technology
at Northwestern University as part of its shared cluster program,
Quest.

\bibliographystyle{apj}
\bibliography{cosmo}

\begin{thebibliography}{64}
\expandafter\ifx\csname natexlab\endcsname\relax\def\natexlab#1{#1}\fi

\bibitem[{{Abel} {et~al.}(2002){Abel}, {Bryan}, \& {Norman}}]{Abel+02}
{Abel}, T., {Bryan}, G.~L., \& {Norman}, M.~L. 2002, Science, 295, 93

\bibitem[{{Barkana} \& {Loeb}(2002)}]{BL02}
{Barkana}, R. \& {Loeb}, A. 2002, \apj, 578, 1

\bibitem[{{Barkana} \& {Loeb}(2005)}]{BL05}
---. 2005, \mnras, 363, L36

\bibitem[{{Barkana} \& {Loeb}(2011)}]{BL11}
---. 2011, \mnras, 839

\bibitem[{{Benson} {et~al.}(2002{\natexlab{a}}){Benson}, {Frenk}, {Lacey},
  {Baugh}, \& {Cole}}]{Benson02a}
{Benson}, A.~J., {Frenk}, C.~S., {Lacey}, C.~G., {Baugh}, C.~M., \& {Cole}, S.
  2002{\natexlab{a}}, \mnras, 333, 177

\bibitem[{{Benson} {et~al.}(2002{\natexlab{b}}){Benson}, {Lacey}, {Baugh},
  {Cole}, \& {Frenk}}]{Benson02b}
{Benson}, A.~J., {Lacey}, C.~G., {Baugh}, C.~M., {Cole}, S., \& {Frenk}, C.~S.
  2002{\natexlab{b}}, \mnras, 333, 156

\bibitem[{{Bittner} \& {Loeb}(2011)}]{Bittner+11}
{Bittner}, J.~M. \& {Loeb}, A. 2011, ArXiv e-prints

\bibitem[{{Bovy} \& {Dvorkin}(2012)}]{BD}
{Bovy}, J. \& {Dvorkin}, C. 2012, ArXiv e-prints

\bibitem[{{Bromm} {et~al.}(1999){Bromm}, {Coppi}, \& {Larson}}]{BCL99}
{Bromm}, V., {Coppi}, P.~S., \& {Larson}, R.~B. 1999, \apjl, 527, L5

\bibitem[{{Bromm} {et~al.}(2002){Bromm}, {Coppi}, \& {Larson}}]{BCL02}
---. 2002, \apj, 564, 23

\bibitem[{{Bromm} \& {Yoshida}(2011)}]{BY11}
{Bromm}, V. \& {Yoshida}, N. 2011, ArXiv e-prints

\bibitem[{{Bullock} {et~al.}(2000){Bullock}, {Kravtsov}, \&
  {Weinberg}}]{Bullock}
{Bullock}, J.~S., {Kravtsov}, A.~V., \& {Weinberg}, D.~H. 2000, \apj, 539, 517

\bibitem[{{Ciardi} {et~al.}(2006){Ciardi}, {Scannapieco}, {Stoehr}, {Ferrara},
  {Iliev}, \& {Shapiro}}]{Ciardi+06}
{Ciardi}, B., {Scannapieco}, E., {Stoehr}, F., {Ferrara}, A., {Iliev}, I.~T.,
  \& {Shapiro}, P.~R. 2006, \mnras, 366, 689

\bibitem[{{Clark} {et~al.}(2011){Clark}, {Glover}, {Klessen}, \&
  {Bromm}}]{Clark11}
{Clark}, P.~C., {Glover}, S.~C.~O., {Klessen}, R.~S., \& {Bromm}, V. 2011,
  \apj, 727, 110

\bibitem[{{Dalal} {et~al.}(2010){Dalal}, {Pen}, \& {Seljak}}]{Dalal+10}
{Dalal}, N., {Pen}, U., \& {Seljak}, U. 2010, \jcap, 11, 7

\bibitem[{{Fialkov} {et~al.}(2011){Fialkov}, {Barkana}, {Tseliakhovich}, \&
  {Hirata}}]{Fialkov+11}
{Fialkov}, A., {Barkana}, R., {Tseliakhovich}, D., \& {Hirata}, C.~M. 2011,
  ArXiv e-prints

\bibitem[{{Furlanetto} \& {Oh}(2006)}]{Furlanetto06}
{Furlanetto}, S.~R. \& {Oh}, S.~P. 2006, \apj, 652, 849

\bibitem[{{Gnedin}(2000)}]{gnedin00}
{Gnedin}, N.~Y. 2000, \apj, 542, 535

\bibitem[{{Gnedin} \& {Hui}(1998)}]{cs}
{Gnedin}, N.~Y. \& {Hui}, L. 1998, \mnras, 296, 44

\bibitem[{{Gnedin} {et~al.}(2008){Gnedin}, {Kravtsov}, \& {Chen}}]{Gnedin+08}
{Gnedin}, N.~Y., {Kravtsov}, A.~V., \& {Chen}, H. 2008, \apj, 672, 765

\bibitem[{{Greif} {et~al.}(2011){Greif}, {White}, {Klessen}, \&
  {Springel}}]{Greif+11}
{Greif}, T., {White}, S., {Klessen}, R., \& {Springel}, V. 2011, ArXiv e-prints

\bibitem[{{Greif} {et~al.}(2010){Greif}, {Glover}, {Bromm}, \&
  {Klessen}}]{Greif10}
{Greif}, T.~H., {Glover}, S.~C.~O., {Bromm}, V., \& {Klessen}, R.~S. 2010,
  \apj, 716, 510

\bibitem[{{Hoeft} {et~al.}(2006){Hoeft}, {Yepes}, {Gottl{\"o}ber}, \&
  {Springel}}]{Hoeft+06}
{Hoeft}, M., {Yepes}, G., {Gottl{\"o}ber}, S., \& {Springel}, V. 2006, \mnras,
  371, 401

\bibitem[{{Iliev} {et~al.}(2003{\natexlab{a}}){Iliev}, {Scannapieco}, {Martel},
  \& {Shapiro}}]{iliev2}
{Iliev}, I.~T., {Scannapieco}, E., {Martel}, H., \& {Shapiro}, P.~R.
  2003{\natexlab{a}}, \mnras, 341, 81

\bibitem[{{Iliev} {et~al.}(2003{\natexlab{b}}){Iliev}, {Scannapieco}, {Martel},
  \& {Shapiro}}]{iliev+03}
---. 2003{\natexlab{b}}, \mnras, 341, 81

\bibitem[{{Iliev} {et~al.}(2005){Iliev}, {Scannapieco}, \& {Shapiro}}]{iss05}
{Iliev}, I.~T., {Scannapieco}, E., \& {Shapiro}, P.~R. 2005, \apj, 624, 491

\bibitem[{{Jeans}(1928)}]{jeans}
{Jeans}, J.~H. 1928, {Astronomy and cosmogony}, ed. {Jeans, J.~H.}

\bibitem[{{Komatsu} {et~al.}(2009){Komatsu}, {Dunkley}, {Nolta}, {Bennett},
  {Gold}, {Hinshaw}, {Jarosik}, {Larson}, {Limon}, {Page}, {Spergel},
  {Halpern}, {Hill}, {Kogut}, {Meyer}, {Tucker}, {Weiland}, {Wollack}, \&
  {Wright}}]{wmap5}
{Komatsu}, E., {Dunkley}, J., {Nolta}, M.~R., {Bennett}, C.~L., {Gold}, B.,
  {Hinshaw}, G., {Jarosik}, N., {Larson}, D., {Limon}, M., {Page}, L.,
  {Spergel}, D.~N., {Halpern}, M., {Hill}, R.~S., {Kogut}, A., {Meyer}, S.~S.,
  {Tucker}, G.~S., {Weiland}, J.~L., {Wollack}, E., \& {Wright}, E.~L. 2009,
  \apjs, 180, 330

\bibitem[{{Kuhlen} {et~al.}(2006){Kuhlen}, {Madau}, \& {Montgomery}}]{Kuhlen}
{Kuhlen}, M., {Madau}, P., \& {Montgomery}, R. 2006, \apjl, 637, L1

\bibitem[{{Maio} {et~al.}(2011){Maio}, {Koopmans}, \& {Ciardi}}]{Maio+11}
{Maio}, U., {Koopmans}, L.~V.~E., \& {Ciardi}, B. 2011, \mnras, L197+

\bibitem[{{McQuinn} {et~al.}(2007){McQuinn}, {Lidz}, {Zahn}, {Dutta},
  {Hernquist}, \& {Zaldarriaga}}]{mcquinn07}
{McQuinn}, M., {Lidz}, A., {Zahn}, O., {Dutta}, S., {Hernquist}, L., \&
  {Zaldarriaga}, M. 2007, \mnras, 377, 1043

\bibitem[{{McQuinn} \& {O'Leary}(2012)}]{McOL12}
{McQuinn}, M. \& {O'Leary}, R.~M. 2012, ArXiv e-prints

\bibitem[{{More} {et~al.}(2011){More}, {Kravtsov}, {Dalal}, \&
  {Gottl{\"o}ber}}]{More+11}
{More}, S., {Kravtsov}, A.~V., {Dalal}, N., \& {Gottl{\"o}ber}, S. 2011, \apjs,
  195, 4

\bibitem[{{Naiman} {et~al.}(2011){Naiman}, {Ramirez-Ruiz}, \&
  {Lin}}]{Naiman+11}
{Naiman}, J.~P., {Ramirez-Ruiz}, E., \& {Lin}, D.~N.~C. 2011, \apj, 735, 25

\bibitem[{{Naoz} \& {Barkana}(2005)}]{NB05}
{Naoz}, S. \& {Barkana}, R. 2005, \mnras, 362, 1047

\bibitem[{{Naoz} \& {Barkana}(2007)}]{NB07}
---. 2007, \mnras, 377, 667

\bibitem[{{Naoz} \& {Barkana}(2008)}]{NB08}
---. 2008, \mnras, 385, L63

\bibitem[{{Naoz} {et~al.}(2009){Naoz}, {Barkana}, \& {Mesinger}}]{NBM}
{Naoz}, S., {Barkana}, R., \& {Mesinger}, A. 2009, \mnras, 399, 369

\bibitem[{{Naoz} {et~al.}(2006){Naoz}, {Noter}, \& {Barkana}}]{NNB}
{Naoz}, S., {Noter}, S., \& {Barkana}, R. 2006, \mnras, 373, L98

\bibitem[{{Naoz} {et~al.}(2011){Naoz}, {Yoshida}, \& {Barkana}}]{Naoz+10}
{Naoz}, S., {Yoshida}, N., \& {Barkana}, R. 2011, \mnras, 416, 232

\bibitem[{{Naoz} {et~al.}(2012){Naoz}, {Yoshida}, \& {Gnedin}}]{Naoz+11a}
{Naoz}, S., {Yoshida}, N., \& {Gnedin}, N.~Y. 2012, \apj, 747, 128

\bibitem[{{Okamoto} {et~al.}(2008){Okamoto}, {Gao}, \& {Theuns}}]{Okamoto+08}
{Okamoto}, T., {Gao}, L., \& {Theuns}, T. 2008, \mnras, 390, 920

\bibitem[{{O'Leary} \& {McQuinn}(2012)}]{OLMc12}
{O'Leary}, R.~M. \& {McQuinn}, M. 2012, ArXiv e-prints

\bibitem[{{Press} \& {Schechter}(1974)}]{ps}
{Press}, W.~H. \& {Schechter}, P. 1974, \apj, 187, 425

\bibitem[{{Ricotti} {et~al.}(2002{\natexlab{a}}){Ricotti}, {Gnedin}, \&
  {Shull}}]{Ric+02}
{Ricotti}, M., {Gnedin}, N.~Y., \& {Shull}, J.~M. 2002{\natexlab{a}}, \apj,
  575, 33

\bibitem[{{Ricotti} {et~al.}(2002{\natexlab{b}}){Ricotti}, {Gnedin}, \&
  {Shull}}]{Ricotti+02}
---. 2002{\natexlab{b}}, \apj, 575, 49

\bibitem[{{Shapiro} {et~al.}(2006){Shapiro}, {Ahn}, {Alvarez}, {Iliev},
  {Martel}, \& {Ryu}}]{Shapiro+06}
{Shapiro}, P.~R., {Ahn}, K., {Alvarez}, M.~A., {Iliev}, I.~T., {Martel}, H., \&
  {Ryu}, D. 2006, \apj, 646, 681

\bibitem[{{Shapiro} \& {Giroux}(1987)}]{shapiro87}
{Shapiro}, P.~R. \& {Giroux}, M.~L. 1987, \apjl, 321, L107

\bibitem[{{Shapiro} {et~al.}(2004){Shapiro}, {Iliev}, \& {Raga}}]{Shapiro+04}
{Shapiro}, P.~R., {Iliev}, I.~T., \& {Raga}, A.~C. 2004, \mnras, 348, 753

\bibitem[{{Somerville}(2002)}]{Somerville}
{Somerville}, R.~S. 2002, \apjl, 572, L23

\bibitem[{{Springel}(2005)}]{G2}
{Springel}, V. 2005, \mnras, 364, 1105

\bibitem[{{Springel} {et~al.}(2001){Springel}, {Yoshida}, \& {White}}]{Gadget}
{Springel}, V., {Yoshida}, N., \& {White}, S.~D.~M. 2001, \na, 6, 79

\bibitem[{{Stacy} {et~al.}(2011){Stacy}, {Bromm}, \& {Loeb}}]{Stacy+10}
{Stacy}, A., {Bromm}, V., \& {Loeb}, A. 2011, \apjl, 730, L1

\bibitem[{{Trenti} {et~al.}(2010){Trenti}, {Smith}, {Hallman}, {Skillman}, \&
  {Shull}}]{Trenti_halos}
{Trenti}, M., {Smith}, B.~D., {Hallman}, E.~J., {Skillman}, S.~W., \& {Shull},
  J.~M. 2010, \apj, 711, 1198

\bibitem[{{Trenti} \& {Stiavelli}(2009)}]{Trenti+09}
{Trenti}, M. \& {Stiavelli}, M. 2009, \apj, 694, 879

\bibitem[{{Tseliakhovich} {et~al.}(2010){Tseliakhovich}, {Barkana}, \&
  {Hirata}}]{Tes+10b}
{Tseliakhovich}, D., {Barkana}, R., \& {Hirata}, C. 2010, ArXiv e-prints

\bibitem[{{Tseliakhovich} \& {Hirata}(2010)}]{Tes+10a}
{Tseliakhovich}, D. \& {Hirata}, C. 2010, \prd, 82, 083520

\bibitem[{{Visbal} {et~al.}(2012){Visbal}, {Barkana}, {Fialkov},
  {Tseliakhovich}, \& {Hirata}}]{Visbal+12}
{Visbal}, E., {Barkana}, R., {Fialkov}, A., {Tseliakhovich}, D., \& {Hirata},
  C. 2012, ArXiv e-prints

\bibitem[{{Yoo} {et~al.}(2011){Yoo}, {Dalal}, \& {Seljak}}]{Yoo+11}
{Yoo}, J., {Dalal}, N., \& {Seljak}, U. 2011, \jcap, 7, 18

\bibitem[{{Yoshida} {et~al.}(2003{\natexlab{a}}){Yoshida}, {Abel}, {Hernquist},
  \& {Sugiyama}}]{Yoshida+03early}
{Yoshida}, N., {Abel}, T., {Hernquist}, L., \& {Sugiyama}, N.
  2003{\natexlab{a}}, \apj, 592, 645

\bibitem[{{Yoshida} {et~al.}(2007){Yoshida}, {Omukai}, \&
  {Hernquist}}]{Yoshida+07}
{Yoshida}, N., {Omukai}, K., \& {Hernquist}, L. 2007, \apjl, 667, L117

\bibitem[{{Yoshida} {et~al.}(2008){Yoshida}, {Omukai}, \&
  {Hernquist}}]{Y08_firstS}
---. 2008, Science, 321, 669

\bibitem[{{Yoshida} {et~al.}(2006){Yoshida}, {Omukai}, {Hernquist}, \&
  {Abel}}]{Yoshida+06}
{Yoshida}, N., {Omukai}, K., {Hernquist}, L., \& {Abel}, T. 2006, \apj, 652, 6

\bibitem[{{Yoshida} {et~al.}(2003{\natexlab{b}}){Yoshida}, {Sugiyama}, \&
  {Hernquist}}]{Yoshida03b}
{Yoshida}, N., {Sugiyama}, N., \& {Hernquist}, L. 2003{\natexlab{b}}, \mnras,
  344, 481

\end{thebibliography}

\appendix

\section{Fit calculation}\label{app:fit}
 
\begin{figure}
  \centering \includegraphics[clip,scale=0.6]{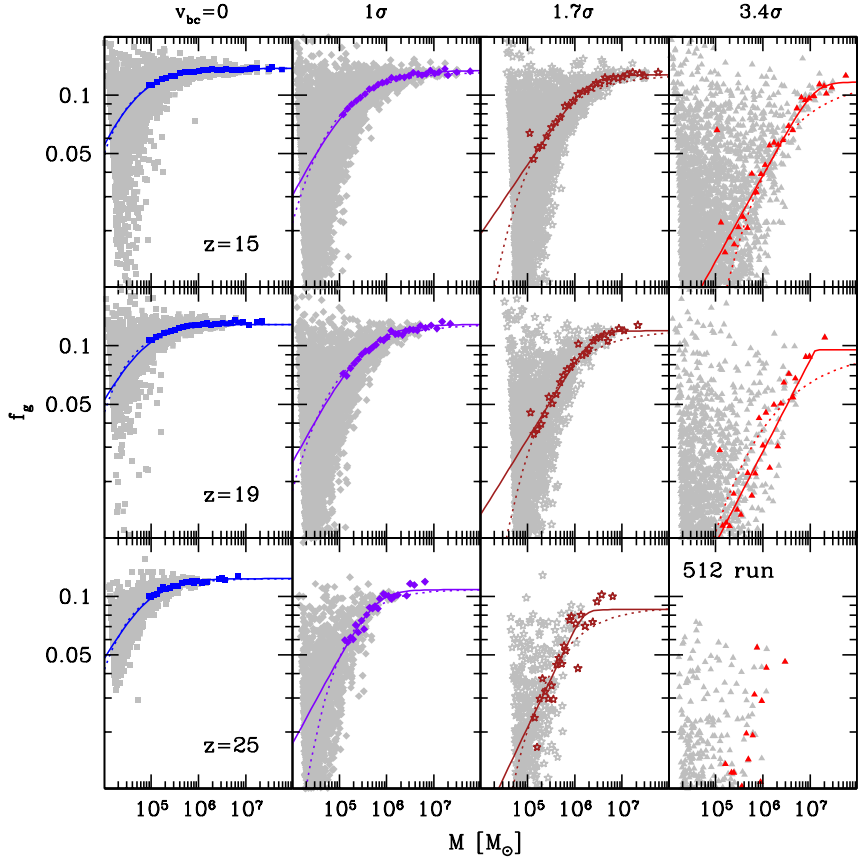}
\caption{A representative example of the gas fraction as a function of mass for the $N=512$ set for three different redshifts (top to bottom) $z=15,19$ and $z=20$. We compare between the new fit model, solid lines,  using eq.~(\ref{f_g-new}), and the old one, dotted line, using eq.~(\ref{f_g-alpha}. We show all of the halo of which $N_h\geq 100$ (grey points) as well as the binned data points for $N_h\geq500$.  We consider the various values of the stream velocity (from left to right).
 $v_\bc=0$, $v_\bc=1\sigma_\vbc$, $v_\bc=1.7\sigma_\vbc$, and
  $v_\bc=3.4\sigma_\vbc$.  Note for the case of $v_\bc=3.4\sigma_\vbc$,  at $z=25$, about $2/3$ of the halos have less than $1 \%$ of gas in them, and $f_{\br,0}$ is very low. Thus no reliable  fit could be found, and therefore we did not show the resulted red line in that case.  } \label{fig:fit_fg}
\end{figure}

\begin{figure}
  \centering \includegraphics[clip,scale=0.6]{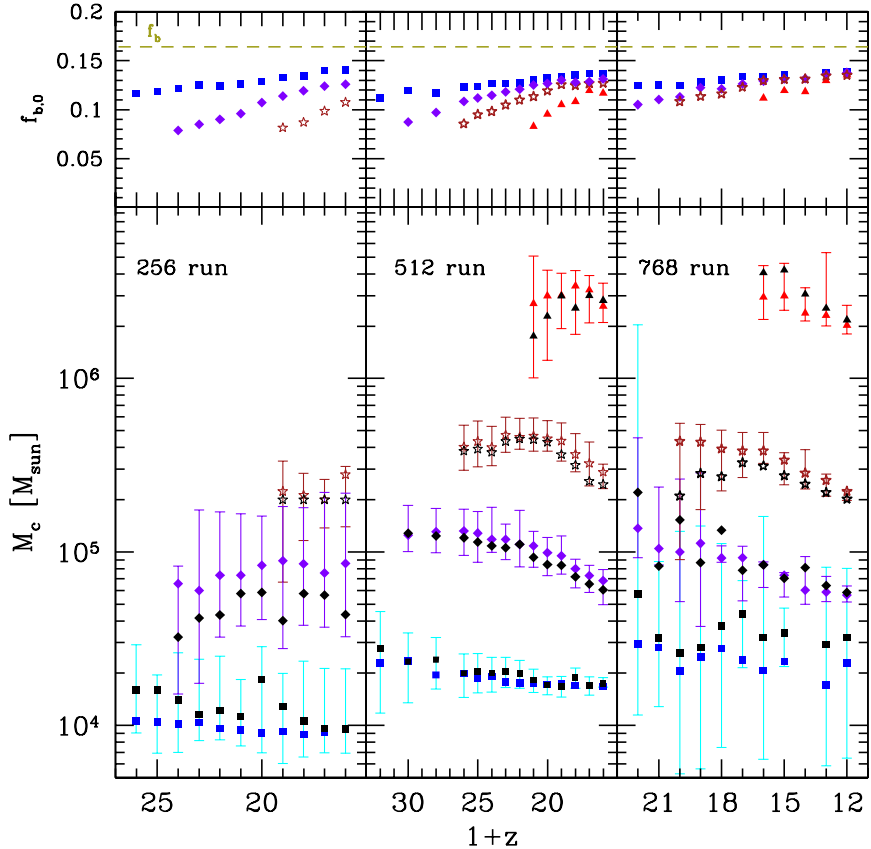}
\caption{Comparison between the new fit model and the old one. We consider in black the old model, using eq.~(\ref{f_g-alpha}), and the color points are  the fit for the new model, using eq.~(\ref{f_g-new}), calculated with minimum of 100 particles per halo (see text). We consider (from left to right) the $N=256,512$ and $N=768$ sets.  We consider the various values of the stream velocity
 $v_\bc=0$, $v_\bc=1\sigma_\vbc$, $v_\bc=1.7\sigma_\vbc$, and
  $v_\bc=3.4\sigma_\vbc$ (blue squares, purple diamonds, brown stars  and red triangles  respectively).  } \label{fig:fit_calc}
\end{figure}

 As can see in Figure \ref{fig:fg512} (dotted lines), the fitted formula found by \citet{gnedin00}, i.e., eq.~(\ref{f_g-alpha}), dose not capture the behavior of the gas fraction as a function of mass for the cases of non-negligible stream velocity. 
In our search for better fitting formula we found that the low mass tail has a significant effect on the fit for large $v_\bc$ values. It is not surprising since the stream velocity deprives the low mass halos of gas at high redshift,  as we showed in Paper I. Therefore, this induce a large dependency on the behavior  of the gas fraction in the low mass tail. 
However, the gas fraction of the low mass halos with less than $500$ particles is poorly constrained  \citep[as was shown in ][]{NBM}.
 Therefore, we introduce a weight function, that account for the errors in estimating the gas fraction as a function of the number of particle in a halo, $N_h$. We use \citet{NBM} resolution study, their figure 4, and assume a scatter of $20 \%$ for halos that have more than $500$ particles. For smaller number of particles per halo we adopted a simply linear function of the error of the gas fraction as can be estimated from \citet{NBM}, figure 4. Thus the   weight function has the following form:
\begin{equation}
\mathcal{W}(N_h)= \left\{ 
\begin{array}{ll} 
 0.2 &  \text{if } N_h\geq 500 \ ,\\
7.5\times 10^{-4} N_h + 0.575 & \text{if } N_h<500 \ .
 \end{array} \right.
\end{equation}
Using this weight function we find the fit, using the forma formula [eq.~(\ref{f_g-alpha})] and the new formula eq.~(\ref{f_g-new}).
In figure \ref{fig:fit_fg} we show an example for the gas fraction as a function of mass, and the two fits as in Figure \ref{fig:fg512} (solid lines for the new formula and dotted lines for the former one). Here we also show all of the points we considered in evaluating the fit (i.e., $N_h\geq 100$). We note that the $\chi^2$ for the new formula fit is for most cases higher (closer to one) than the $\chi^2$ of the old formula, both lower than unity.

The ultimate goal of the fitting process is to find the characteristic mass $M_c$. An important test is to compare the resulted $M_c$ from the two fitting formulae. 
The different values achieved for the two models are shown in Figure \ref{fig:fit_calc}. As depicted in this Figure, the $M_c$ for the different values of $v_\bc$ converge over the different ranges of redshift for the $N=512$ and $N=768$ sets. However, or the $N=256$ set the 
new formula  produces a systematically low value for $M_c$ even for the $v_\bc=0$ case. As mentioned in Paper I, this run suffers from low statistic, particularly in the large mass tail,. Furthermore, there was no convergence of the gas fraction, these are the main cause for this systematics.  Therefore, based on the $N=512$ and $N=768$ sets, we conclude that although the new formula produces somewhat better fit, the final $M_c$ results did not changed by much. This is not surprising  since the meaning of the two models is the same.

As shown in  Figure  \ref{fig:fit_fg}, the new formula results in a plateau at the  high mass tail, which indicates a possible degeneracy between the fitting parameters. This of course present a problem in evaluating the values and errors of $M_c$. However, as shown in Figure \ref{fig:fit_calc} the values of $M_c$ from the two fitting models are consistent.
Therefore, in evaluating the errors of the fitting values of $M_c$ we choose to be conservative and select the larger values between the  $1-\sigma$ errors from the two models. 
In addition we have used bootstrap method for the new model, in some cases of the $N=512$ set to test our evaluation of the errors\footnote{We could not use the bootstrap method to all of the runs since in few runs we have somewhat low statistic sample.  }, and found that they are consistent with choosing the maximum $1-\sigma$ from the two fit models.  We show the complete best fit parameters for  the $512$ set in table \ref{tablefit512}, they are similar for the other sets. Note that for high redshift and large $v_\bc$ values the best fit parameters are poorly constrain. Specifically the parameters  $\beta$ and $\gamma$ from equation (\ref{f_g-new}) are sometimes so poorly  constrains (i.e., more then an order of magnitude) that we omit the errors from the table, thus the symbol ``$--$" in table \ref{tablefit512} means error larger then an order of magnitude, for those cases $\chi^2$ was  close to zero.

\begin{figure}
  \centering \includegraphics[clip,scale=0.6]{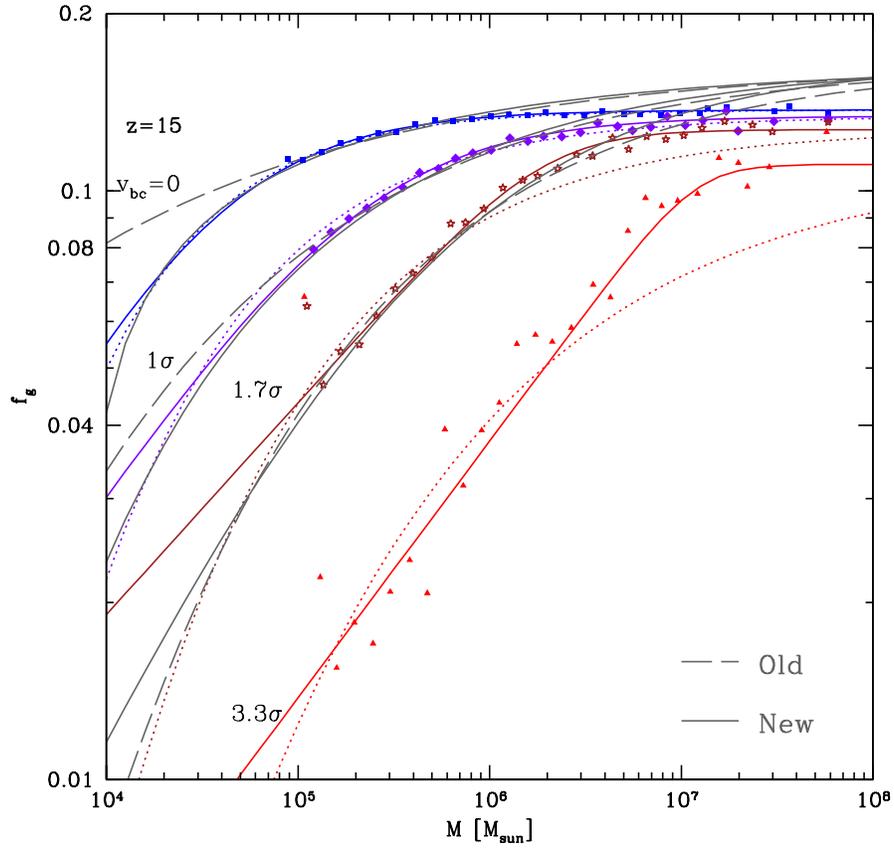}
\caption{Comparison between the new fit model and the old one while using $f_{\br,0}$ and $\bar{f}_{\br}$. We consider the  $N=512$ run for $z=15$ representative example. We compare between using $\bar{f}_\br$ value, solid (dotted) grey lines for the new (old) model. We also show the fit while using the $f_{\br,0}$ value, solid (dotted) color lines for the new (old) model.
 We consider the various values of the stream velocity
 $v_\bc=0$, $v_\bc=1\sigma_\vbc$, $v_\bc=1.7\sigma_\vbc$, and
  $v_\bc=3.4\sigma_\vbc$ (blue squares, purple diamonds, brown stars  and red triangles  respectively).  } \label{fig:fg_fbbar}
\end{figure}

Note that in our calculation of the fit we used $f_{\br,0}$ which is the gas fraction in the high mass tail. This value is lower than the mean cosmic baryonic fraction $\bar{f}_\br$, since the baryons are lagging behind the dark matter even at high redshifts \citep[see][]{NB07,BL11}. As shown in  \citet[][fig.~4]{Naoz+10},  this causes  a higher  $f_{\br,0}$  for the initial conditions that assumes $\delta_\br=\delta_\dm$ than the $f_{\br,0}$ resulted from the smother baryonic initial conditions, as explored in their  other two initial conditions models\footnote{In both cases $f_{\br,0}$ is defined in the same way as described here.}.
In addition to the physical reason, a numerical reason may arise, since some baryons are in fact bound to the halo and we miss them due to our  halo finder algorithm (\S \ref{sec:Hdef}). As can be seen in Figure  \ref{fig:fg_fbbar} assuming  $f_{\br,0}\to \bar{f}_\br$, grey lines, by either of the models results in a worse fit than using the calculated $f_{\br,0}$.

 \begin{table}
 \caption{ The best-fit parameters from equation~(\ref{f_g-new}) for the ${N=512}$ set.}
\label{tablefit512}
\begin{center}
\begin{tabular}{l l l l l}
\hline
Redshift & $M_c$   [M$_\odot$]              &  $\beta$  & $\gamma$\\
\hline \hline
  $v_\bc=0$& &   &\\
\hline 
$31$ & $2.8_{-1.1}^{+2.2}\times 10^4$&  $  1.45 \pm 0.71 $&$  3.4  \pm 0.8$\\
$29$ & $2.3_{-1.}^{+1.1}\times 10^4$&  $  1.09 \pm 0.1 $&$  2.33  \pm 0.5$\\
$27$ & $2_{-0.3}^{+1.3} \times 10^4$&  $ 1.37\pm  0.1$&$ 3.37\pm    0.5 $\\
$25$   & $2_{-0.5}^{+0.6}\times 10^4$&  $ 1.14\pm 0.06$&$   2.68\pm  0.33$\\
$24$ & $1.9_{-0.4}^{+0.7}\times 10^4$& $1.15\pm 0.06$&$    2.76\pm   0.31   $\\
$23$   & $1.8_{-0.4}^{+0.6}\times 10^4$ &  $1.04\pm   0.04$&$ 2.16\pm    0.25$\\
$22$   & $1.8_{-0.1}^{+0.7}\times 10^4$&  $1.17\pm 0.05 $&$  2.87\pm    0.26 $  \\
$21$   & $1.7_{-0.1}^{+0.6}\times 10^4$&  $ 1.15\pm  0.04$&$    2.71\pm 0.23 $  \\
$20$   & $1.7_{-0.2}^{+0.4}\times 10^4$&  $1.05\pm  0.03$&$ 2.35\pm    0.2$\\
$19$   & $1.7_{-0.2}^{+0.2}\times 10^4$ &  $ 0.96\pm 0.03$&$1.91  \pm  0.18   $ \\
$18$   & $1.7_{-0.3}^{+0.2}\times 10^4$ &  $ 0.91\pm 0.02$&$ 1.66\pm 0.15  $\\
$17$   & $1.7_{-0.7}^{+0.4}\times 10^4$&  $0.87\pm0.02$&$1.43 \pm 0.15    $\\
$16$   & $1.7_{-0.2}^{+0.2}\times 10^4$ &  $ 0.88\pm  0.02 $&$ 1.36  \pm 0.14  $ \\
$15$   & $1.7_{-0.2}^{+0.2}\times 10^4$ &  $0.91\pm 0.02$&$ 1.47 \pm   0.14 $\\
    \hline 
 $v_\bc=1\sigma_\vbc$  & &  &\\
    \hline 
 $29$ & $1.2_{-0.2}^{+0.6}\times 10^5$&  $  7.35\pm   11.88$&$15.78\pm   25.79$\\
$27$ & $1.3_{-0.3}^{+0.5} \times 10^5$&  $ 6.13\pm   5.86$&$  14.14\pm   13.78$\\
$25$   & $1.3_{-0.3}^{+0.4}\times 10^5$&  $ 1.75\pm 0.33$&$ 3.89 \pm  0.91$\\
$24$ & $1.3_{-0.4}^{+0.4}\times 10^5$& $1.64\pm    0.25$&$    3.55\pm   0.69    $\\
$23$   & $1.2_{-0.2}^{+0.6}\times 10^5$ &  $1.44 0.19$&$3.2\pm   0.57$\\
$22$   & $1.2_{-0.3}^{+0.3}\times 10^5$&  $ 1.38\pm    0.15$&$  3.1\pm    0.47$  \\
$21$   & $1.1_{-0.3}^{+0.6}\times 10^5$&  $  1.21\pm   0.11$&$ 2.7\pm 0.38$  \\
$20$   & $1.1_{-0.3}^{+0.2}\times 10^5$&  $  1.12 \pm  0.09$&$  2.52\pm    0.32$\\
$19$   & $9.9_{-2.6}^{+2.2}\times 10^4$ &  $  1.01\pm0.07$&$2.19    \pm0.28  $ \\
$18$   & $9.5_{-2}^{+2.8}\times 10^4$ &  $   0.97\pm 0.06$&$ 2.17  \pm 0.26 $\\
$17$   & $8.1_{-1.8}^{+1.1}\times 10^4$&  $ 1.05\pm  0.06 $&$  2.4 \pm 0.26    $\\
$16$   & $7.2_{-1.4}^{+1.1}\times 10^4$ &  $ 1.01 0.05$&$  2.22 \pm  0.23 $ \\
$15$   & $6.8_{-1.8}^{+1.1}\times 10^4$ &  $ 0.97 0.05$&$ 2.15 \pm  0.22 $\\
\hline 
$v_\bc=1.7\sigma_\vbc$& &     &\\
\hline 
$25$   & $4_{-1.1}^{+1.3}\times 10^5$&  $4.45\pm   6.74$&$    8.7\pm  13.68 $\\
$24$ & $4.3_{-1.2}^{+1.3}\times 10^5$& $3.94 \pm  5.16 $&$  8.19\pm11.15    $\\
$23$   & $4_{-0.9}^{+1.3}\times 10^5$ &  $ 3.58  \pm 3.14$&$ 7.36 \pm  6.8$\\
$22$   & $4.7_{-1}^{+1.3}\times 10^5$&  $  4.96\pm 5.98 $&$ 11.03  \pm 13.63$  \\
$21$   & $4.6_{-0.7}^{+1.3}\times 10^5$&  $  2.91  2$&$ 6.89 \pm   5.05$  \\
$20$   & $4.6_{-0.9}^{+1.2}\times 10^5$&  $   3.45\pm  2.35$&$   8.5 \pm  6.06$\\
$19$   & $4.5_{-0.7}^{+1.2}\times 10^5$ &  $  2.64 \pm1.23  $&$  6.7 \pm  3.36 $ \\
$18$   & $4.3_{-1}^{+1.2}\times 10^5$ &  $   2.06\pm  0.69$&$ 5.26  \pm 1.97 $\\
$17$   & $3.6_{-0.7}^{+1.2}\times 10^5$&  $ 2.07\pm  0.7$&$ 5.69  \pm  2.16  $\\
$16$   & $3.2_{-0.8}^{+1.1}\times 10^5$ &  $   1.75\pm    0.41$&$ 4.7  \pm  1.31 $ \\
$15$   & $2.9_{-0.6}^{+0.3}\times 10^5$ &  $  1.77\pm 0.38$&$  4.9\pm   1.25 $\\
\hline 
$v_\bc=3.4\sigma_\vbc$& &     &\\
\hline
$20$   & $2.7_{-1.7}^{+2.3}\times 10^6$&  $   102 --$&$  222 --$\\
$19$   & $3_{-1.7}^{+1.2}\times 10^6$ &  $  88 --  $&$  187 --$ \\
$18$   & $3_{-1.1}^{+1}\times 10^6$ &  $   4.82\pm  13.1$&$ 10.3  --$\\
$17$   & $3.4_{-1.6}^{+0.7}\times 10^6$&  $ 3.56\pm  7.9$&$ 8.16 \pm  18.4  $\\
$16$   & $3.2_{-1.4}^{+0.7}\times 10^6$ &  $   2.1\pm    1.23$&$ 4.8  \pm  2.9 $ \\
$15$   & $2.6_{-0.5}^{+0.9}\times 10^6$ &  $  2.33\pm 1.35$&$  5.3\pm   3.2 $\\
\hline 
\vspace{-0.7cm}
\end{tabular}
\end{center}
\end{table}

\end{document}